\documentclass[3p, times, onecolumn, review]{elsarticle}
\usepackage{lineno,hyperref}

\usepackage{soul}
\usepackage[usenames,dvipsnames,svgnames,table]{xcolor} 										
\setstcolor{red}
\usepackage{hyperref}
\usepackage[hyphenbreaks]{breakurl}

\journal{\ \ }

\newcommand{\ve}[1]{\textbf{\textit{#1}}}

\begin{document}

\begin{frontmatter}

\title{Effects of electromagnetic fluctuations in plasmas on solar neutrino fluxes }

\author[SSU]{Eunseok Hwang}

\author[CoReLS]{Dukjae Jang\corref{mycorrespondingauthor}}
\cortext[mycorrespondingauthor]{Corresponding author}
\ead{djjang2@ibs.re.kr}

\author[SSU]{Kiwan Park}

\author[Beihang,NAOJ]{Motohiko Kusakabe}

\author[Beihang,NAOJ,UT]{Toshitaka Kajino} 

\author[UWM,NAOJ]{A. Baha Balantekin}

\author[NU,NAOJ]{Tomoyuki Maruyama}

\author[SSU]{Youngshin Kwon}

\author[UNIST]{Kyujin Kwak}

\author[SSU,Beihang,NAOJ]{Myung-Ki Cheoun}

\address[SSU]{Department of Physics and OMEG Institute, Soongsil University, Seoul 156-743, Republic of Korea}
\address[CoReLS]{Center for Relativistic Laser Science, Institute for Basic Science (IBS), Gwangju 61005, Republic of Korea}
\address[Beihang]{School of Physics and International Research Center for Big-Bang Cosmology and Element Genesis, Beihang University, Beijing 100083, China}
\address[NAOJ]{National Astronomical Observatory of Japan, 2-21-1 Osawa, Mitaka, Tokyo 181-8588, Japan}
\address[UT]{The University of Tokyo, Bunkyo-ku, Tokyo 113-0033, Japan}
\address[UWM]{Physics Department, University of Wisconsin-Madison,1150 University Avenue, Madison, Wisconsin 53706, USA}
\address[NU]{College of Bioresource Sciences, Nihon University, Fujisawa 252-0880, Kanagawa-ken, Japan}
\address[UNIST]{Department of Physics, Ulsan National Institute of Science and Technology (UNIST), Ulsan 44919, Republic of Korea}

\begin{abstract}
We explore the effects of electromagnetic (EM) fluctuations in plasmas on solar neutrino fluxes exploiting the fluctuation-dissipation theorem. We find that the EM spectrum in the solar core is enhanced by the EM fluctuations due to the high density of the Sun, which increases the radiation energy density and pressure. By the EM fluctuations involving the modified radiation formula, the central temperature decreases when the central pressure of the Sun is fixed. With a help of the empirical relation between central temperature and neutrino fluxes deduced from the numerical solar models, we present the change in each of the solar neutrino fluxes by the EM fluctuations. We also discuss the enhanced radiation pressure and energy density by the EM fluctuations for other astronomical objects.
\end{abstract}


\end{frontmatter}


\section{Introduction}
 Astrophysical plasmas in nucleosynthesis sites have been generally presumed to be ideal, implying that the thermonuclear reaction rate is determined by equilibrium velocity distribution and nuclear reaction cross sections for bare nuclei. However, the collective motions and collisions of constituent particles in astrophysical plasmas could affect the electromagnetic interactions in the nucleosynthesis, which motivated studies of impact of astrophysical plasmas on nucleosynthesis yields. A typical example is electrons near an ion screen the nuclear charge enhancing the nuclear reaction rates \cite{salpeter1954}. Such screening effects on thermonuclear reaction rates in plasmas have been widely discussed for the solar interior \cite{carraro1988, bahcall2002, adelberger2011} and the early universe \cite{itoh_1997, itoh2002, wang2011, famiano2016, luo2020, Hwang2021}. Also, studies on big bang nucleosynthesis (BBN) with Tsallis distribution function involving soft energy spectra \cite{tsallis1988} have suggested a partial solution to the primordial lithium problem \cite{Bertulani2013, Hou2017, Kusakabe2019}. A solution has been proposed by a transient model of the photon distribution function during BBN, which has speculated that the transition is related to the plasma properties \cite{jang2021}.

In this paper, we mainly discuss another notable phenomenon in a plasma: the electromagnetic (EM) fluctuations. Thermal fluctuations exist even in a homogeneous plasma maintaining the thermal (or nearly thermal) equilibrium. In other words, the root mean square of EM fields may manifest themselves even without external fields. The level of EM fluctuations can be evaluated by the fluctuation-dissipation theorem \cite{sitenko1967}. Pioneering works have studied the EM fluctuations near the zero-frequency in cold and warm plasmas \cite{tajima_pop1992, cable1992, opher_magnetic_1997}, and subsequently one has discussed the implication of EM fluctuations on the origin of cosmological magnetic fields \cite{tajima_apj1992}, radiation spectrum in the BBN epoch \cite{opher1997}, and nuclear reaction rates by highly damped modes in the stellar interior \cite{opher2001}.

Since the effect of EM fluctuations is more dominant under the high density and low temperature conditions, we expect that the modification of the radiation spectrum by EM fluctuations becomes significant in the core of the Sun rather than in the early universe studied previously in the literature \cite{tajima_pop1992, cable1992, opher_magnetic_1997}. Exploiting the fluctuation-dissipation theorem, we find that the radiation spectrum is significantly enhanced in the solar core which has higher density. This implies that the energy of photons in the solar core could be larger at a given temperature than the blackbody spectrum, which affects the pressure and energy density of the radiation in the solar core where Planck distribution is adopted in the standard treatment. In the core, if the pressure is kept as a constraint, the enhanced EM spectrum causes a decrease in the temperature. On the other hand, when the temperature is retained, the total pressure including the EM fluctuations is higher than the pressure in the standard solar model (SSM).

If the plasma temperature changes due to the EM fluctuations, it affects the nuclear reaction rates in the Sun, and also the production of the solar neutrinos. Then, the solar neutrino fluxes, especially for the uncertain CNO neutrino fluxes, would differ from the prediction in the SSM. In this paper, adopting the empirical relation between the solar neutrino fluxes and the central temperature of the SSM \cite{bahcall1996}, we estimate the feasible change in the solar neutrino fluxes resulting from the changed central temperature affected by the EM fluctuations. The result based on the empirical relation would provide a guidance to study the EM fluctuation effects on the Sun, although a rigorous study requires solving the stellar structure equations. We also discuss the effects of EM fluctuations on other astronomical objects and provide the modified radiation pressure in a wide parameter space of density and temperature.

The rest of this paper is organized as follows. In section 2, we introduce the formalism of the fluctuation-dissipation theorem for EM fields and show the distorted EM spectrum in the solar core. In section 3, using the EM spectrum, we derive a formula for the modified radiation pressure in the Sun. Then, we present the changes in central temperature and solar neutrino fluxes. Finally, in section 4, we discuss the effects of the EM fluctuations on stellar evolution and summarize our conclusion. We adopt the natural units for all equations in this paper, i.e., $\hbar =c \equiv 1$.

\section{Electromagnetic spectrum by fluctuations}
From the fluctuation-dissipation theorem, the fluctuations of magnetic ($\ve{B}$) and transverse electric (${\ve E}_{\rm T}$) fields are derived as \citep{sitenko1967}, 
\begin{eqnarray}
\frac{\left\langle {\ve B}^2 \right\rangle_{{\ve k}, \omega}}{8\pi} &=& \frac{2}{\exp[\omega/T]-1} \left( \frac{k}{\omega} \right)^2 \frac{{\rm Im}\,\epsilon_{\rm T}(\omega, {\ve k})}{\left| \epsilon_{\rm T} (\omega, {\ve k}) - \left( \frac{k}{\omega} \right)^2 \right|^2}, \label{eq1} \\ 
\frac{\left\langle {\ve E}_{\rm T}^2 \right\rangle_{{\ve k}, \omega}}{8\pi} &=& \frac{2}{\exp[\omega/T]-1}  \frac{{\rm Im}\,\epsilon_{\rm T}(\omega, {\ve k})}{\left| \epsilon_{\rm T}(\omega, {\ve k}) - \left( \frac{k}{\omega} \right)^2 \right|^2}
\label{eq2},
\end{eqnarray}
where $\omega$, ${\ve k}$, and $T$ denote the angular frequency, wavevector, and temperature, respectively. The transverse dielectric permittivity $\epsilon_{\rm T}(\omega,{\ve k})$ is obtained from the first order Vlasov equation with Bhatnagar-Gross-Krook (BGK) collision term, which is given as \cite{opher_magnetic_1997}
\begin{eqnarray} \label{eq3}
\epsilon_{\rm T} (\omega, {\ve k}) = 1 + \sum_\alpha \frac{\omega_{p\alpha}^2}{\omega^2} \left( \frac{\omega}{\sqrt{2} k v_\alpha} \right) Z\left( \frac{\omega + i\eta_\alpha}{\sqrt{2} k v_\alpha} \right),
\end{eqnarray} 
where $\omega_{p\alpha}$, $\eta_\alpha$, and $v_\alpha$ are the plasma frequency, collision rate, and thermal velocity of species $\alpha$, respectively. We adopt the typical collision frequency for an electron given as $\eta_e = 4\sqrt{2 \pi} n_e e^4 \ln \Lambda / (3 m_e^{1/2} T^{3/2})$, where $n_e$, $m_e$, and $\ln \Lambda$ are the number density and mass of electron, and the Coulomb logarithm taken as $\ln \Lambda = 17$, respectively. We do not consider the collision frequency for the ion as it is inversely proportional to the mass of the ion. The plasma dispersion function also known as Fried-Conte function $Z(z)$ is defined as\footnote{We note that references \cite{opher_magnetic_1997, opher1997} wrote the $e^{t^2}$ instead of $e^{-t^2}$, which is a typo.} \cite{fried_2015}
\begin{eqnarray} \label{eq4}
Z(z) = \frac{1}{\sqrt{\pi}}\int_{-\infty}^\infty \frac{e^{-t^2}}{t-z} dt.
\end{eqnarray}
In fact, a more precise approach requires the Boltzmann collision term involving the integration of all relevant collisions over the momentum space. However, for simplicity, we adopt the BGK collision term as used in Refs.\,\cite{tajima_pop1992, opher_magnetic_1997, tajima_apj1992, opher1997}.

Integrating Eqs.\,(\ref{eq1}) and (\ref{eq2}) over ${\ve k}$, the EM spectrum is obtained as follows:
\begin{eqnarray}
S(\omega) = \int \left[ \frac{\left\langle {\ve B}^2 \right\rangle_{{\ve k} \omega}}{8\pi} + \frac{\left\langle {\ve E}_T^2 \right\rangle_{{\ve k} \omega}}{8\pi} \right] d{\ve k}.
\label{eq5}
\end{eqnarray}
As shown in Eq.\,(\ref{eq3}), for $\omega \gg \omega_{p\alpha}$, the real and imaginary parts of $\epsilon_T({\ve k}, \omega)$ goes to unity and zero, respectively. For this condition, the terms of the dielectric permittivity in Eqs.\,(\ref{eq1}) and (\ref{eq2}) give a Dirac-delta function of the argument $\omega^2 - k^2c^2$ for high frequency. Therefore, the EM spectrum at the high frequency region sustains the blackbody shape satisfying the dispersion relation in free space. In contrast, at the low frequency region of $\omega \lesssim \omega_{p\alpha}$, a change in the transverse dielectric permittivity causes a distortion of EM spectrum depending on the plasma temperature and density. In other words, the collective effects of plasma are significant for a low frequency mode with long wavelength.

For $\omega \rightarrow 0$, the integration in Eq.\,(\ref{eq5}) diverges, and a cutoff parameter for $k$ is required. Although it was argued that the intensity of the magnetic fluctuations is not sensitive for the cutoff \cite{tajima_pop1992, tajima_apj1992}, Opher and Opher disputed the argument providing the sensitivity of the cutoff parameter for the intensity \cite{opher_magnetic_1997}. Instead, they proposed the new cutoff parameter as the inverse distance of closest approach between a test particle and an electron in a plasma, $k_{\rm max} = 1/r_{\rm min} \simeq Mm_ev^2/((M+m_e)|eq|)$ where $M$, $v$, and $q$ indicate the mass, velocity, and charge of the test particle, respectively \cite{ichimaru_2018}. By taking this cutoff parameter $k_{\rm max}$ (or $r_{\rm min}$), one can avoid Coulomb energy of the particles exceeding their kinetic energy by excluding very small distance where the Coulomb interaction is stronger than the kinetic energy. This guarantees the perturbation expansion by the plasma parameter as well as convergence of $S(\omega \rightarrow 0)$. In this paper, we adopt the $k_{\rm max} = Mm_ev^2/((M+m_e)|eq|)$ used in Ref.\,\cite{opher_magnetic_1997} as the upper bound of integration in Eq.\,(\ref{eq5}).

\begin{figure}[h]
\centering
\includegraphics[width=8.5cm]{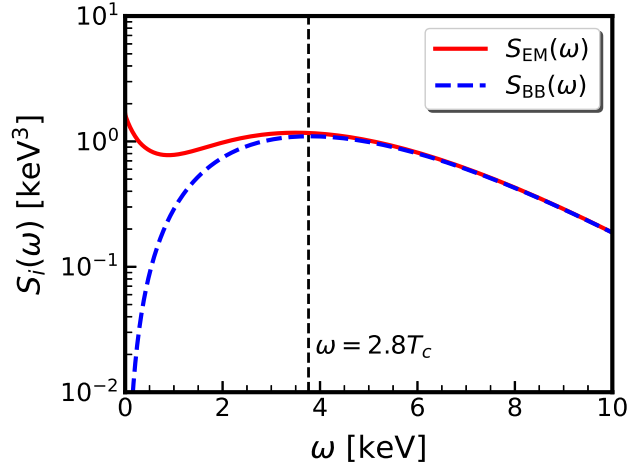}
\caption{EM spectrum in the solar center. Red-solid and blue-dashed lines denote $S_{\rm EM}(\omega)$ and $S_{\rm BB}(\omega)$, respectively. The vertical dashed line indicates the peak intensity region of $\omega = 2.8 T_c$ where $T_c$ denoted the central temperature of the Sun. The density and temperature in the center is adopted as $\rho_c = 6.38 \times 10^{8}\,{\rm keV}^4\ (= 1.48 \times 10^2\,{\rm g/cm^3}\ {\rm in\ cgs\ unit)}$, and $T_c=1.34\,{\rm keV}\ (=1.56\times10^7\,{\rm K})$, respectively \cite{bahcall1988}. Nuclear mass fractions are taken from the SSM \cite{bahcall1988}.}
\label{fig1}
\end{figure}

We computed $S(\omega)$ under the central condition of the Sun, indicating the density and temperature as $\rho_c = 6.38 \times 10^{8}\,{\rm keV}^4\ (= 1.48 \times 10^2\,{\rm g/cm^3}\ {\rm in\ cgs\ unit)}$, $T_c=1.34\,{\rm keV}\ (=1.56\times10^7\,{\rm K})$, respectively \cite{bahcall1988}. Figure \ref{fig1} shows EM spectrum at the given condition. We emphasize that the change in the EM spectrum occurs even in the peak region of the blackbody spectrum, i.e., $\omega_{peak} \approx 2.8 T_c$.  By integrating the EM spectrum over the energy, we obtain that the radiation energy density ($U$) and pressure ($P$) are 1.22 times higher than the blackbody case. Compared to the 1\% increment of radiation energy density at the BBN epoch \cite{opher1997}, the change of radiation energy in solar core is remarkable. 

\section{Effects of EM fluctuations on solar neutrino fluxes}
The change in the radiation spectrum due to EM fluctuations affects the temperature in the solar core if the total pressure, which is the summation of gas and radiation pressure, is kept the same in a fluctuating plasma. Because the EM fluctuations are caused by the electromagnetic interaction between plasma (or gas/matter) and radiation, it is possible in principle that not only the radiation spectrum but also the gas pressure changes due to EM fluctuations. Here we assume that the total pressure does not change (at least significantly) due to EM fluctuations. This assumption needs to be verified further using the first principle approaches, for example, by explicitly solving the (Boltzmann) equations which deal with the radiation-matter interaction. However, our assumption is approximately valid for the following two reasons. First, the total pressure near the core of the sun is dominated by the gas pressure. Second, the modification of the radiation pressure due to EM fluctuations is not large. For the same reasons, we can further assume that the gas and the radiation share the same temperature even if both pressures are modified due to EM fluctuations. 

In the SSM, the total pressure $(P_{\rm SSM})$ is written as follows:
\begin{eqnarray}
P_{\rm SSM} = P_{\rm gas} + P_{\rm rad}^{\rm BB} = \beta P_{\rm SSM} + (1-\beta)P_{\rm SSM},
\label{eq6}
\end{eqnarray}
where $P_{\rm gas}$ and $P_{\rm rad}^{BB}$ are gas and blackbody radiation pressure, respectively, and $\beta$ is the ratio of the gas to the total pressure in the SSM, i.e., $\beta \equiv P_{\rm gas}/P_{\rm SSM}$. With the EM fluctuations, the total pressure $P$ is modified as 
\begin{eqnarray}
P = P_{\rm SSM} + \Delta P_{\rm rad} = P_{\rm gas} + P_{\rm rad}^{\rm BB} + \Delta P_{\rm rad},
\label{eq7}
\end{eqnarray}
where $\Delta P_{\rm rad} (= P^{\rm EM}_{\rm rad} - P^{\rm BB}_{\rm rad})$ denotes the difference in the radiation pressure caused by EM fluctuations. The blackbody radiation pressure is given as $P^{\rm BB}_{\rm rad}=aT^4/3$ by following the Planck distribution where $a$ is the radiation constant. We assume that $P^{\rm EM}_{\rm rad} = a_{\rm EM}(T, \rho, X_i)T^4/3$ for the distorted photon distribution due to EM fluctuations, where $a_{\rm EM}(T, \rho, X_i)$ is the modified radiation constant which depends on the temperature, density, and ion fractions, $X_i$. Then, by defining a new ratio of gas pressure to the total pressure, $\beta^* (=P_{\rm gas}/P)$, Eq.\,(\ref{eq7}) becomes 
\begin{eqnarray}
P = \beta^* P + (1-\beta^*)P,
\label{eq8}
\end{eqnarray}
where $\beta^*$ can be expressed in terms of $\beta$ in SSM as follows:
\begin{eqnarray}
\beta^* = \frac{P_{\rm gas}}{P} = \frac{P_{\rm gas}}{P_{\rm SSM} + \Delta P_{\rm rad}} = \frac{\beta}{1 + \left[a_{\rm EM}(T,\rho,X_i)/a- 1\right] (1-\beta)}.
\label{eq9}
\end{eqnarray}
We note that $\beta^* \simeq 1$ because $\beta \simeq 1$.

In principle, the modified radiation formula due to the EM fluctuations must be evaluated in the entire stellar interior by solving the full set of stellar structure equations. However, since $a(T,\rho, X_i)$ converges to $a$ at low density, the change in the radiation coefficient is small in the outer region. Also, considering that the neutrino fluxes in our interest are sensitive only to the central temperature of the Sun, the effects of the EM fluctuations matter only in the central region as well.

Following our assumption that the total pressure does not change while the gas and the radiation share the same temperature even in the presence of  EM fluctuations, $P_{\rm c,SSM}(T_{\rm c, SSM}) = P_{\rm c,EM}(T_{\rm c, EM})$ we write at the center of the Sun, 
\begin{eqnarray}
\frac{aT_{\rm c,SSM}^4}{3(1-\beta_c)} = \frac{a_{\rm EM}\left(T_{c, \rm EM}, \rho_{c}, X_{i,\rm c} \right)T_{c, \rm EM}^4}{3(1-\beta^*_c)},
\label{eq10}
\end{eqnarray} 
where sub-index $c$ stands for the central condition of the Sun. Substituting Eq.\,(\ref{eq9}) into Eq.\,(\ref{eq10}), we can derive the following relation:
\begin{eqnarray}
\frac{T_{\rm c, SSM}}{T_{c, \rm EM}} = \left[ 1 + \left( \frac{a_{\rm EM}(T_c, \rho_c, X_{i,c})}{a} -1 \right) (1-\beta_c) \right]^{1/4}.
\label{eq11}
\end{eqnarray}
Then, taking $1-\beta_c = 6.52 \times 10^{-4}$ \cite{bahcall1988} and $a_{\rm EM}(T_{c, \rm EM}, \rho_{c}, X_{i,c})/a = 1.22$, we obtain the following ratio of temperature between SSM ($T_{\rm c,SSM}$) and EM fluctuation ($T_{\rm c,EM}$) models:
\begin{eqnarray}
\frac{T_{\rm c,SSM}}{T_{\rm c,EM}} - 1 = 3.7 \times 10^{-5},
\label{eq12}
\end{eqnarray}
which implies that the EM fluctuation in the center of the Sun leads to $3.7\times 10^{-3}$\% decrease of the central temperature in comparison with SSM. Such a reduction of the central temperature would affect the solar neutrino fluxes by changing the thermonuclear reaction rates.

Empirically, in the solar models, it is known that the neutrino fluxes obey the scaling law for the central temperature as follows \cite{bahcall1988}:
\begin{eqnarray}
\phi_i \propto T_c^{m_i},
\label{eq13}
\end{eqnarray}
where $\phi_i$ is the neutrino fluxes produced due to a reaction channel $i$, and $m_i$ denotes a corresponding exponent. Therefore, from Eq.\,(\ref{eq13}), the change of neutrino fluxes due to EM fluctuations is given as 
\begin{eqnarray}
\frac{\phi_{i, \rm EM}}{\phi_{i, \rm SSM}} = \left( \frac{T_{c, \rm EM}}{T_{c, \rm SSM}} \right)^{m_i}.
\label{eq14}
\end{eqnarray} 
Taking the scaling exponent $m_i$ obtained from the 1000 numerical solar models \cite{bahcall1996}, we list the fractional change of neutrino flux for each reaction channel in Tab.\,\ref{table1}. We note that $\Delta \phi <0$ with $m_i < 0$. The negative exponent in the scaling law in Eq.\,(\ref{eq13}) indicates that the neutrino flux increases as the central temperature of the sun decreases. Because the central temperature decreases due to EM fluctuations, the neutrino flux for the reaction channel having a negative exponent such as pp and pep decreases, i.e., $\Delta \phi$ for these two channels are negative. In contrast, $\Delta \phi >0$ for all the other channels which have positive $m_i$.

\begin{table}[htb]
\caption{The fractional change of neutrino flux ($\Delta \phi_i / \phi_{i, \rm SSM}$) for each reaction channel due to EM fluctuations, where $\Delta \phi_i \equiv \phi_{i,\rm SSM} - \phi_{i, \rm EM}$. Exponent $m_i$ for the numerical solar models (second column) and uncertainties in SSM (last three columns) are taken from  \cite{bahcall1996} and \cite{orebigann2021}, respectively.}
   \label{table1}
    \centering
       \begin{tabular}{c c c ccc}
       \hline
       Type of neutrino & 
       Exponent &
       $\Delta \phi_i / \phi_{i, \rm SSM}$ &
       \multicolumn{3}{c}{Uncertainties (\%)}     \\
       \cline{4-6}
        fluxe &   $m_i$  & (\%)  & Nuclear & Environment & CNO\\
       \hline
      $\phi_{pp}$            & -1.1 &  $-4.03 \times 10^{-3}$  &  0.4  & 0.4 & 0.1 \\ \hline
       $\phi_{pep}$          & -2.4 &  $-8.80 \times 10^{-3}$  &  0.6  & 0.8 & 0.3 \\ \hline
       $\phi_{^7{\rm Be}}$   & 10   &  $0.037               $  &  5    & 4.1 & 0.8 \\ \hline
       $\phi_{^8{\rm B}}$    & 24   &  $0.088               $  &  7.6  & 9.2 & 1.9 \\ \hline
       $\phi_{^{13}{\rm N}}$ & 24.4 &  $0.089               $  &  6.2  & 6.9 & 12  \\ \hline
       $\phi_{^{15}{\rm O}}$ & 27.1 &  $0.099               $  &  8.7  & 8.4 & 12  \\ \hline
       $\phi_{^{17}{\rm F}}$ & 27.8 &  $0.102               $  &  9.3  & 9.0 & 16  \\ \hline
       \end{tabular}
\end{table}

Table \ref{table1} includes the current uncertainties in the solar neutrino fluxes estimated in the SSMs. The uncertainties come from three origins \cite{orebigann2021, Vinyoles_2017}. The uncertainty in the nuclear reactions (4th column in Tab.\,\ref{table1}) stems from the uncertainties in the nuclear cross sections, which affect each neutrino flux by changing the relevant nuclear reaction rates. At the central temperature of the sun, it is very challenging to experimentally measure the nuclear cross sections which determine the relevant reaction rates. For this reason, most of (low-energy) cross sections are extrapolated from those measured at higher energies by using the R-matrix theory \cite{descouvemont_2004} and/or the Bayesian analysis \cite{gomez_2017}. The uncertainty in environment (5th column) is caused by the uncertainty in the stellar structure. For example, the temperature profile inside the sun is strongly affected by opacity which is not measured accurately with experiments but relies on theoretical models \citep{iglesias_1996, badnell_2005, colgan_2016}. The uncertainties in CNO (6th column) is due to uncertainties in the CNO metallicity, i.e., the abundance of these elements. Conventionally, solar metallicity is estimated observationally from solar spectra and/or meteorites. The advances in the measurement skills enable solar metallicity to be continually updated, but this update can also be a source of uncertainty, especially for the CNO metallicity. For example, the core (or internal metallicity) on which the neutrino flux depends may be significantly different from the conventional metallicity estimated from the solar spectra which contain the information on the abundances at the solar atmosphere.

The change of neutrino fluxes due to EM fluctuations shown in Tab.\,\ref{table1} does not exceed the current uncertainties estimated in the SSMs. This implies that the effects of EM fluctuations on the solar neutrino fluxes cannot be identified observationally at the moment. For example, modification of the CNO neutrino fluxes due to EM fluctuations is at the level of about 0.1\%, which is not distinguishable in the current uncertainties. However, it is possible to pin down the effects of EM fluctuations when the uncertainties in the other three sources decrease below the level of 0.1\%. Observational measurements of the CNO neutrinos (e.g., by the Borexino collaboration \cite{borexino2020}) and solar metallicity as well as experimental measurements of low-energy cross sections of nuclear reactions and opacities will be able to reduce the current uncertainties in the future.

Another strong constraint to the solar neutrino fluxes is provided by the solar luminosity $L$, and we evaluate whether the changes in neutrino fluxes due to EM fluctuations are consistent with the current error of the solar luminosity. Since PP chains predominantly contribute to the energy production in the Sun (about 99\%), the change in solar luminosity due to EM fluctuations can be approximated with the change in nuclear reaction rates which depend on temperature. Using the obtained decrease of the central temperature in Eq.\,(\ref{eq12}), we obtain the following change in $L$ as
\begin{eqnarray}
\frac{L_{\rm EM}}{L_{\rm SSM}} \approx \frac{R_{33}(T_{\rm c, EM}) + R_{34}(T_{\rm c, EM})}{R_{33}(T_{\rm c, SSM}) + R_{34}(T_{\rm c, SSM})}.
\label{eq16}
\end{eqnarray}
Here we use the fact that the solar luminosity $L$ is dominated by two reactions, ${^3{\rm He}} + {^3{\rm He}} \rightarrow {^4{\rm He}} + 2p\ (33)$ in PP I and ${^3{\rm He}} + {^4{\rm He}} \rightarrow {^7{\rm Be}} + \gamma\ (34)$ in PP II or PP III. These two reactions are often represented as (33) and (34), respectively, and $R_{ij}$ in the above equation indicates the reaction rate of ($ij$). In fact, $L$ is proportional to $\epsilon_{33} R_{33} + \epsilon_{34} R_{34}$, where $\epsilon_{ij}$ is the energy generated in ($ij$). Because $\epsilon_{33} \approx \epsilon_{34}$ and $R_{ij}$ is a function of temperature, we can obtain the above equation. For $R_{33}$ and $R_{34}$, we adopt the JINA REACLIB database \citep{cyburt2010}. We find that the solar luminosity decreases by 0.1\% due to the effects of EM fluctuations which lower the central temperature. This result can be compared with the total energy production including the uncertainty of neutrino fluxes in the SSM, \citep{orebigann2021}
\begin{eqnarray}
L_{\rm SSM} = 1.04^{+0.07}_{-0.08} L_{\odot},
\label{eq17}
\end{eqnarray} 
where $L_{\odot}$ is the measured solar luminosity including measurement errors and the uncertainty in Eq.\,(\ref{eq17}) results from the neutrino fluxes. Again, the effects of EM fluctuations on the solar luminosity are not differentiated under the current uncertainties in the SSM, even though those effects are not considered as uncertainty sources in the SSM. However, one may consider the effect of EM fluctuations as one of the energy loss processes. For example, other non-standard solar models which exploit exotic particles like axion \citep{graham_2015, sikivie_2021} have been proposed to reduce the overestimated central value in Eq.\,(\ref{eq17}). Future measurements of the solar luminosity could validate the non-standard models if the measurement errors were similar to the predicted errors of the non-standard models.

Among the three uncertainty sources in the SSM, we can separately estimate the uncertainties in nuclear reactions. The uncertainty in the solar luminosity solely based upon the current experimental results (i.e., uncertainties in nuclear reaction cross sections) is expressed as  
\begin{eqnarray}
L_{\rm LC} = (0.991^{+0.005}_{-0.005} + 0.009^{+0.004}_{-0.005} )L_{\odot},
\label{eq18}
\end{eqnarray}  
where $L_{\odot}$ is the observed solar luminosity including measurement errors as in Eq.\,(\ref{eq17}) and LC stands for luminosity constraint as in \cite{orebigann2021}. The first and second terms imply the contributions by the $pp$ chain and CNO cycle, respectively. For the $pp$ chain, as mentioned above, the $L_{\rm EM}$ estimated by reactions of (33) and (34) decreases due to EM fluctuations by 0.1\,\%, which is also allowed within the uncertainty in Eq.\,(\ref{eq18}). Furthermore, the contribution of CNO cycles could be evaluated using the reaction rates at $^{14}{\rm N}(p,\gamma)^{15}{\rm O}$ that determines the rate of the cycle as the slowest reaction. Comparing the reaction rate for $T_{\rm c,SSM}$ and $T_{\rm c,EM}$, we obtain a 0.12\% decrease of the $^{14}{\rm N}(p,\gamma)^{15}{\rm O}$ at the center of the Sun, which is rather comparable to the uncertainty of the CNO part in $L_{\rm LC}$. Therefore, a more accurate computation considering the entire  stellar structure with the EM fluctuation effect may be needed if experimental uncertainties decrease.

\section{Conclusion and discussion}
In conclusion, based on the fluctuation-dissipation theorem, we presented the effects of EM fluctuations on the center of the Sun. The EM fluctuations in the solar core enhance the EM spectrum and increase the radiation pressure for the given temperature. Assuming a fixed pressure in the center of the Sun, we have shown that the central temperature decreases by $3.7 \times 10^{-3}$\% due to EM fluctuations, and tabulated the change of solar neutrino fluxes due to the EM fluctuations in Tab.\,\ref{table1}. In particular, we showed that effects of EM fluctuations could reduce the total energy production of the Sun by about 0.1\,\%. This implies that the effects of EM fluctuations could be considered as one of the energy loss processes in the solar core within the uncertainties in SSMs. However, verification of this change is too challenging a task given the limits of  current observations. We expect that future experiments with greater precisions would verify the effect of the EM fluctuations on solar neutrino fluxes. 

Finally, in addition to the effects on solar neutrinos, we discuss several issues concerning the effects of EM fluctuations on stellar evolution as follows.

First, the energy generated by the nuclear fusions in the stellar interior is transferred to the surface by three mechanisms: conduction, convection, and radiative diffusion. Among those processes, radiative diffusion is dominant in transporting the energy in the inner region, depending on the radiation property. In SSMs, the radiation energy density is given by the Planck law, but it would increase by an additional spectrum if the EM fluctuation due to the plasma is considered. Therefore, in the stellar interior, the heat flux could be changed by the additional radiation energy density, which would affect the radiative energy transport. Such effects of additional radiation energy density should be estimated and examined to be compatible with the measured solar luminosity.

Second, although we only consider the central pressure and temperature in this paper, EM fluctuations could occur at each point in stellar interior, affecting the stellar profiles of density, temperature, and luminosity. Since the density rapidly decreases from the core to surface, the significant change in such profiles by EM fluctuations would appear only in the inner region. Through the investigation of EM fluctuation effects on stellar profiles, one can evaluate the validity of EM fluctuations using other observables such as surface temperature, luminosity, and sound velocity. However, precise investigation for the stellar evolution requires consistent computation to solve the requisite set of equations: mass conservation, hydrostatic equilibrium, energy transport, and energy generation. We leave a detailed analysis treating the stellar structure to a future work.

\begin{figure}[h] 
\centering
\includegraphics[width=8.5cm]{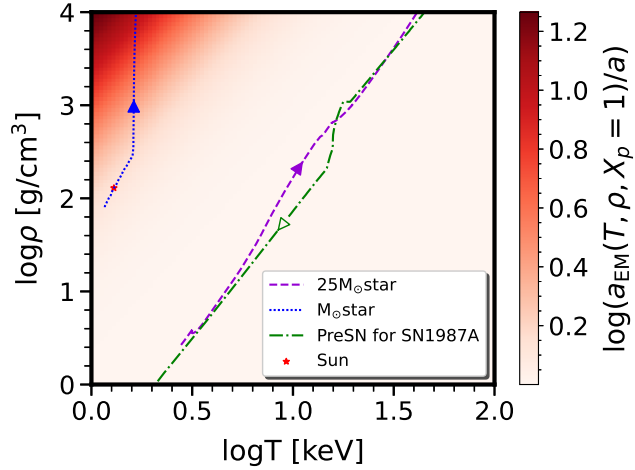}
\caption{The ratio of pressure due to EM fluctuations to blackbody formula on the parameter plane of the temperature and density. The blue dotted, purple dashed, and green-dashed-dotted lines indicate profiles of $1M_{\odot}$ \citep{christensen_2021}, and $25M_\odot$ \citep{paxton_2011, paxton_2015, paxton_2018, paxton_2019} stars, and preSN for SN1987A \citep{nomoto_1984, shigeyama_1990}, respectively. The filled arrows denote the time evolution of the profiles, and unfilled arrow presents the PreSN profile for SN1987A from inner to outer regions before the explosion. The red star symbol denotes the core condition of the Sun in the SSM.}
\label{fig2}
\end{figure}

Finally, we consider the effects of EM fluctuations on other astronomical objects rather than the Sun. Figure \ref{fig2} shows the ratio of radiation coefficients by the EM fluctuation to the standard blackbody value in the parameter space of the temperature and density for electron-proton plasma. As shown in Fig.\,\ref{fig2}, the radiation pressure is significantly affected by EM fluctuations only above a critical density for a given temperature, and the critical density increases with increasing temperature. This implies that EM fluctuations may become significant for higher density and lower temperature region, and affect not only the present solar condition but the evolution of other stars. For further application, we also provide the following analytical fitting formula of $a_{\rm EM}(\rho, T, X_p=1)/a$ in the range of $1 \le T \le 50$ (in unit of keV) and $1 \le \rho \le 10^4$ (in unit of ${\rm g/cm^3}$):
\begin{eqnarray}
\frac{a_{\rm EM}(\rho, T, X_p=1)}{a} = 1 + a(T) \rho - b(T) \rho^{4/3}
\end{eqnarray}
with coefficients $a(T)$ and $b(T)$ given as
\begin{eqnarray}
a(T) &=& \exp\left[ a_0  +  a_1 T^{1/3}  + a_2 T^{2/3} + a_3 T + a_4 T^{4/3} + a_5 T^{5/3} \right], \label{eq_a} \\[12pt]
b(T) &=& \exp\left[ b_0 +  b_1 T^{1/3}  + b_2 T^{2/3} + b_3 T + b_4 T^{4/3} + b_5 T^{5/3} \right] \label{eq_b},
\end{eqnarray} 
where each coefficient is tabulated in Tab.\,2. Ultimately, it would be desirable if theoretical calculations of the stellar evolution as well as neutrino fluxes in the future is based on the stellar codes with the exact radiation pressure including EM fluctuations. 
\begin{table}[htb]
\caption{Fitting coefficients for $a(T)$ and $b(T)$ in Eqs.\,(\ref{eq_a}) and (\ref{eq_b}).}
   \label{table2}
    \centering
       \begin{tabular}{c|c|c|c|c|c}
       \hline
        $a_0$   &   $a_1$   &   $a_2$   &   $a_3$   &   $a_4$   &   $a_5$ \\
       \hline
        $-2.149 \times 10$ & $6.890 \times 10^2 $ & $-9.520 \times 10^3$ &  $5.670 \times 10^4$ & $-1.573 \times 10^5$ & $1.633 \times 10^5$ \\ \hline \hline
       $b_0$    & $b_1$     &   $b_2$   & $b_3$     & $b_4$     & $b_5$ \\
       \hline
       $-3.875 \times 10$   & $1.139 \times 10^3 $ & $-1.479 \times 10^4$ &  $8.512 \times 10^4$ & $-2.340 \times 10^5$ & $2.450 \times 10^5$ \\ 
       \hline
       \end{tabular}
\end{table}

\section*{Acknowledgement}
The work of D.J. is supported by Institute for Basic Science under IBS-R012-D1. E.H., K.P., Y.K. and M.K.C. are supported by the National Research Foundation of Korea (Grant Nos. NRF-2021R1A6A1A03043957 and NRF-2020R1A2C3006177). K.P. is also supported by NRF-2021R1I1A1A01057517 and Y.K. is supported by the National Research Foundation of Korea (NRF) grant funded by the Korea goverment (MSIT) (No. 2019R1A2C1087107). The work of M.K. was supported by NSFC Research Fund (12175010). T.K. is supported in part by Grants-in-Aid for Scientific Research of JSPS (20K03958, 17K05459) and T.M. is supported by JSPS (19K03833). A.B.B. is supported in part by the U.S. National Science Foundation grant No. PHY-2108339 and acknowledges support from the NAOJ visiting professor program. K.K is supported by the National Research Foundation of Korea (Grant Nos. NRF-2016R1A5A1013277 and NRF-2022R1F1A1073890).


\bibliography{./bib/EM.bib}

\end{document}